\documentstyle[sprocl]{article}

\def\Journal#1#2#3#4{{#1} {\bf #2}, #3 (#4)}

\def\PRD{{\em Phys. Rev.} D}

\def\CQG{\em Class. Quantum Gravity}

\def\be{\begin{equation}}
\def\ee{\end{equation}}
\def\bea{\begin{eqnarray}}
\def\eea{\end{eqnarray}}

\begin{document}

\title{GRAVITATIONAL COLLAPSE, BLACK HOLES AND \\ NAKED SINGULARITIES}

\author{T. P. SINGH}

\address{Tata Institute of Fundamental Research, 
\\Homi Bhabha Road, Mumbai 400 005, India\\E-mail: tpsingh@tifrc3.tifr.res.in}

\maketitle\abstracts{This article gives an elementary review of gravitational 
collapse and the cosmic censorship hypothesis. Known models of collapse 
resulting in the formation of black holes and naked singularities are 
summarized. These models, when taken together, suggest that the censorship 
hypothesis may not hold in classical general relativity. The nature of the
quantum processes that take place near a naked singularity, and their
possible implication for observations, is briefly discussed.}

\section{Introduction}

After a star has exhausted its nuclear fuel, it can no longer remain in
equilibrium and must ultimately undergo gravitational collapse. The star
will end as a white dwarf if the mass of the collapsing core is less than
the famous Chandrashekhar limit of 1.4 solar masses. It will end as a
neutron star if the core has a mass greater than the Chandrashekhar limit
and less than about 3-5 times the mass of the sun. It is often believed that
a core heavier than about 5 solar masses will end, not as a white dwarf or
as a neutron star, but as a black hole. However, this belief that a black hole
will necessarily form is not based on any firm theoretical evidence. An
alternate possibility allowed by the theory is that a naked singularity can
form, and the purpose of the present article is to review our current
understanding of gravitational collapse and the formation of black holes and
naked singularities.

A black hole has been appropriately described by Chandrashekhar as the most
beautiful macroscopic object known to man. Only a few parameters suffice to
describe the most general black hole solution, and these objects have
remarkable thermodynamic properties. Further, excellent observational
evidence for their existence has developed over the years~\cite{rees}. 
Thus, there can
be no doubt about the reality of black holes, and the gravitational collapse
of very many sufficiently massive stars must end in the formation of a black
hole.

However, the following question is still very much open. If the collapsing
core is heavy enough to not end as a neutron star, does this guarantee that
a black hole will necessarily form? The answer to this question has to come
from the general theory of relativity, and unfortunately this remains an
unsolved problem.

What we do know from general relativity about gravitational collapse is
broadly contained in the celebrated singularity theorems of Geroch, Hawking
and Penrose. It has been shown that under fairly general conditions, a
sufficiently massive collapsing object will undergo continual gravitational
collapse, resulting in the formation of a gravitational singularity. The
energy density of the collapsing matter, as well as the curvature of
spacetime, are expected to diverge at this singularity.

Is such a singularity necessarily surrounded by an invisible region of
spacetime, i.e. has a black hole formed? The singularity theorems do not
imply so. The singularity may or may not be visible to a far away observer.
If the singularity is invisible to a far away observer, we say the star has
ended as a black hole. If it is visible, we say the star has ended as a
naked singularity. We need to have a better understanding of general
relativity in order to decide whether collapse always ends in a black hole
or whether naked singularities can sometimes form.

Given this situation, Penrose was led to ask~\cite{pe} whether there might 
exist a
cosmic censor who forbids the existence of naked singularities, `clothing
each one of them with a horizon'? Later, this led to the cosmic censorship
hypothesis, which in broad physical terms states that the generic
singularities arising in the gravitational collapse of physically reasonable
matter are not naked. Till today, this hypothesis remains unproven in
general relativity, neither is it clear that the hypothesis holds true in
the theory. What is of course true is that the hypothesis forms the working
basis for all of black hole physics and astrophysics. If cosmic censorship
were to not hold, then some of the very massive stars will end as black
holes, while others could end as naked singularities. As we will argue in
Section 3, these two kinds of objects have very different observational
properties.

There are various very important reasons for investigating whether or not
cosmic censorship holds in classical general relativity. As we have
mentioned above, the hypothesis is vital for black hole astrophysics.
Unfortunately this fact is rarely appreciated by the astrophysics community.
The hypothesis is also necessary for the proof of the black hole area
theorem. It is not clear what the status of this theorem will be if the
hypothesis were to not hold. If naked singularities do occur in classical
relativity, they represent a breakdown of predictability, because one could
not predict the evolution of spacetime beyond a naked singularity. Such
singularities would then provide pointers towards a modification of
classical general relativity, so that a suitable form of predictability is
restored in the modified theory. Further, naked singularities might be
observable in nature, if they are allowed by general relativity. Undoubtedly
then, it is important to find out if the censorship hypothesis is valid.

We wish to make two further remarks. Firstly, while a great deal is known
about the properties of stationary black holes, we know very little
about the process of black hole {\it formation}. In fact we know as little
about the formation of black holes as we do about the formation of naked
singularities. Secondly, it has sometimes been remarked that a theory of
quantum gravity is likely to get rid of the singularities of classical
general relativity, irrespective of whether these singularities are naked or
covered. Why then does it eventually matter whether or not cosmic censorship
holds? The answer to this legitimate objection is the following. A quantum
gravity theory is expected to smear out a classical singularity and replace
it by a region of very high, albeit finite, curvature. If the classical
singularity is hidden behind a horizon (i.e. is a black hole), this quantum
smeared region remains invisible to an external observer. However, if the
classical singularity is naked, the smeared region of very high curvature
will be visible to far away observers, and the physical processes taking
place near this smeared region will be significantly different from those
taking place outside the horizon of ordinary astrophysical black holes.
Hence, from such an experimental standpoint, quantum gravity has little
bearing on the question of cosmic censorship. To put it differently, quantum
gravity is not expected to restore the event horizon, if the horizon
is absent in the classical theory.

Since a theorem proving or disproving the hypothesis has not been found,
attention has shifted to studying model examples of gravitational collapse,
to find out whether the collapse ends in a black hole or a naked
singularity. While specialised examples such as have been studied are
nowhere near a general proof, they are really all that we have to go by, as
of now. However, there does seem to be an underlying pattern in the results
that have been found in these examples, which gives some indication of the
general picture. It is interesting that all models studied to date admit
both black hole and naked singularity solutions, depending on the choice of
initial data. In the next section, we give a summary of what has been learnt
from these examples and what they probably tell us about cosmic censorship.
In the third section we will address the question of whether naked
singularities might occur in nature, and if so, what they would look like to
an observer.  

The reader is also invited to study a few other excellent 
reviews~\cite{revi} on the
subject of cosmic censorship, which have appeared in recent years. Some of
these reviews emphasize aspects other than those presented here, and some
arrive at conclusions on cosmic censorship other than those given here. I
would also like to draw attention to an earlier review of mine on this topic,
which is somewhat more detailed, though a little dated~\cite{tps}. 
Also, for a detailed
bibliography the reader is requested is look up this earlier review; the 
references in the present article are not exhaustive, and largely confined
to the more recent papers.

\section{Theoretical Evidence for the Formation of Black Holes and
Naked Singularities}

We consider the gravitational collapse of physically reasonable classical
matter, where by `physically reasonable' is meant that the matter satisfies
one or more of the energy conditions (weak energy condition, strong energy
condition and the dominant energy condition). Also, most of the collapse
studies that have been carried out so far deal with spherical collapse -
even this simplest of systems is poorly understood, in so far as cosmic
censorship is concerned. It is of course true that spherical collapse, if
allowed to proceed to completion, results in a Schwarzschild black hole.
However, a spherical collapsing system can also admit timelike or null
singularities which can be naked. From the point of view of an observer
falling with the star this can happen if a singularity forms inside the
star (say at its center) before the boundary of the star enters its 
Schwarzschild radius. Such a singularity can be naked.

Since the exact solution of Einstein equations for spherical collapse       
with a general form of matter is not known, collapse of matter with various
equations of state has been studied. In the following pages, we review some
of these results.

\subsection{Spherical Dust Collapse}

Historically, the earliest model of gravitational collapse is due to
Oppenheimer and Snyder. They showed that the collapse of a homogeneous
dust sphere results in the formation of a black hole (by dust is meant
an idealized perfect fluid for which the pressure is zero). It was thought
that the formation of the black hole will not be affected even if the 
specialized assumptions of this model (homogeneity, sphericity, dust equation
of state) are relaxed. However, we now know that this is not so.   

When the assumption of homogeneity is relaxed, there is an exact solution
of Einstein equations - the Datt-Tolman-Bondi solution, which describes
collapse of a dust sphere with non-uniform initial density. Two kinds of
singularities can result - shell crossing and shell focusing. While the
former have a Newtonian analog, at least some of the latter appear to be
of purely relativistic origin. It has been shown by various 
authors~\cite{dust} that
the shell-focusing singularities can be of both the black hole and naked
type, depending on the initial conditions.

As an illustration, we mention the following interesting case. Consider the
collapse of a dust sphere, starting from rest, and having an initial density
profile near the center given by
$$  \rho (R) = \rho_{0}  +  \rho_{1}R  +  {1\over 2} \rho_{2} R^{2}
   +  {1\over 6} \rho_{3} R^{3} + ....   $$
It turns out that the singularity is naked if $\rho_{1}$ is less than zero,
and also if $\rho_{1}$ is equal to zero and $\rho_{2}$ is less than zero. If
both $\rho_{1}$ and $\rho_{2}$ are zero and $\rho_{3}$ is negative, then we
define a dimensionless quantity $\xi=|\rho_{3}|/\rho_{0}^{5/2}$. The 
singularity is naked if $\xi\geq 25.48$ and covered if $\xi$ is less than
this number. If $\rho_{1}$, $\rho_{2}$ and $\rho_{3}$ are all zero, the
singularity is covered, the Oppenheimer-Snyder collapse being a special case
of this.  

It is known that the collapse of null dust (directed radiation), described
by the Vaidya spacetime, also gives rise to both black hole and naked
singularity solutions, depending on the rate of infall~\cite{vaidya}.

\subsection{Spherical Collapse of Fluids with Pressure}      

Exact solutions of Einstein equations describing collapse of fluids are
rare. Hence little is known about the end state of collapse in these systems.
To some degree, numerical methods have been used to integrate Einstein    
equations and study light propagation. In comoving coordinates, the 
energy-momentum tensor is diagonal and its components are the energy density,
the radial pressure and the tangential pressure. For a perfect fluid, the
two pressures are identical.

A significant development was the work of Ori and Piran~\cite{op}, who 
investigated the self-similar gravitational collapse of a perfect fluid
with an equation of state $p=k\rho$. It is readily shown that the collapse
leads to the formation of a curvature singularity. The assumption of 
self-similarity reduces Einstein equations to ordinary differential 
equations which are solved numerically, along with the equations 
for radial and non-radial null geodesics. It is then shown that for every 
value of $k$ (in the range investigated: $0\leq k\leq 0.4$) there are
solutions with a naked singularity, as well as black hole solutions.

An analytical treatment for this problem was developed by Joshi and
Dwivedi~\cite{dj}. After deriving the Einstein equations for the collapsing
self-similar perfect fluid they reduce the geodesic equation, in the
neighborhood of the singularity, to an algebraic equation. The free
parameters in this algebraic equation are in principle determined by
the initial data. The singularity will be naked for those values of the
parameters for which this equation admits positive real roots. Since
this is an algebraic equation, it will necessarily have positive roots
for some of the values of the parameters, and for the initial data
corresponding to such values of the parameters the singularity is
naked. 

Lifshitz and Khalatnikov~\cite{lk} (and Podurets~\cite{po}) worked out the 
form of the solution
near the singularity for the equation of state of radiation. This work
is a precursor to the Belinskii-Lifshitz-Khalatnikov (BLK) series solutions 
near singularities. Following the method of Podurets, we investigated the
nature of the non-central shell-focusing singularity which can form
during the collapse of a fluid~\cite{coo}. It is easily shown that such a 
singularity
is covered so long as the radial pressure is positive. By considering the
case of a perfect fluid, we showed that negative pressure allows for a naked
singularity if the ratio of the pressure to the density is $\leq -1/3$; and
the singularity is covered if this ratio exceeds $-1/3$. We note that the
weak energy condition allows for pressure to be negative, although it is
questionable whether negative pressures could develop during the final
stages of realistic stellar collapse. The BLK series solutions offer a
promising avenue for investigating cosmic censorship, which deserves to be
pursued further.

On physical grounds, imperfect fluids are more realistic than perfect
ones; very little is known about their collapse properties though. An
interesting paper is the one by Szekeres and Iyer~\cite{si}, who do not 
start by assuming an equation of state. Instead they assume the metric
components to have a certain power-law form, and also assume that collapse
of physically reasonable fluids can be described by such metrics. The
singularities resulting in the evolution are analysed, with attention
being concentrated on shell-focusing singularities at $r>0$. They find that
although naked singularities can occur, this requires that the radial or
tangential pressure must either be negative or equal in magnitude to the
density. 

Another model which has recently attracted some attention is the collapse
of a fluid having only tangential pressure~\cite{tp}. The analysis of the 
Einstein
equations is considerably simpler than the case in which radial pressure
is also present. Hence this is a useful system for studying the stability
of dust naked singularities against the introduction of pressure. It has
been found that while certain equations of state admit only black hole
type singularities, other state equations admit naked singularities as well.

We conclude this brief discussion of fluids by commenting on the issue of
whether fluids are a reasonable form of matter in so far as cosmic
censorship studies are concerned. An objection sometimes raised against
fluids is that they form singularities even in Minkowski spacetime, and hence
the naked singularities that have been found do not have anything to do with
general
relativity. However, while some of the singularities, like the shell-crossings,
and possibly the weak shell-focusings, have Minkowskian analogs, it is by
no means clear that all the singularities (for instance the strong curvature
shell-focusings) have counterparts in flat spacetime evolution. At the very
least, this has to be investigated further, so as to get a precise distinction
of the singularities that have Minkowskian analogs, from those that
are of purely relativistic origin. 

It is however more useful to examine cosmic censorship keeping the 
astrophysical context in mind, and here we know that a fluid description of
stellar matter is physically quite appropriate. Thus, if a naked singularity
were to result in the collapse of a real star made of fluid matter, we
would be compelled to seriously pursue its observational consequences.

\subsection{Collapse of a Massless Scalar Field}

In a series of papers, Christodoulou~\cite{ch} has pioneered analytical 
studies of
the spherical collapse of a self-gravitating massless scalar field. He 
established the 
global existence and uniqueness of solutions for the collapsing field, and 
also gave sufficient conditions
for the formation of a trapped surface. For a self-similar scalar collapse
he showed that there are initial conditions which result in the formation
of naked singularities. 

Christodoulou was also interested in the question of the mass of the
black hole which might form during the collapse of the scalar wave-packet.
Given a one parameter family ${\cal S}[p]$ of solutions labeled by the 
parameter $p$
which controls the strength of interaction, it was expected that as $p$
is varied, there would be solutions with $p\rightarrow p_{weak}$ in which
the collapsing wave-packet disperses again, and solutions with
$p\rightarrow p_{strong}$ which have black hole formation. For a given
family there was expected to be a critical value $p=p_{*}$ for which the
first black hole appears as $p$ varies from the weak to the strong range.
Do the smallest mass black holes have finite or infinitesimal mass?
This issue would be of interest for censorship, since an infinitesimal
mass would mean one could probe arbitrarily close to the singularity.

This problem was studied by Choptuik~\cite{cho} numerically and some remarkable
results were found. He confirmed that the family ${\cal S}[p]$
has dispersive solutions as well as those forming black holes, and a 
transition takes place from one class to the other at a critical $p=p_{*}$.
The smallest black holes have infinitesimal mass. Near the critical
region, the mass $M_{bh}$ of the black hole scales as 
$M_{bh}\approx (p-p_{*})^{\gamma}$, where $\gamma$ is a universal constant
(i.e. same for all families) having a value of about 0.37. The near critical
evolution can be described by a universal solution of the field equations
which also has a periodicity property called echoing, or discrete 
self-similarity. That is, it remains unchanged under a rescaling
$(r,t)\rightarrow (e^{-n\Delta}r, e^{-n\Delta}t)$ of spacetime coordinates.
$n$ is an integer, and $\Delta$ is about 3.4. Subsequently, these results
have been confirmed by others.

At the critical solution, the mass of the forming black hole goes to zero,
as $p\rightarrow p_{*}$ from the right. This critical solution is a
naked singularity. However, since the naked singularity is realised
for a specific solution in the one parameter family, it is a subset of
measure zero. As regards cosmic censorship, the more significant feature
are the black holes of arbitrarily small mass, and hence arbitrarily high
curvature that is visible to a far away observer. It is more physical to
think of censorship in terms of whether or not regions of unbounded high
curvature are generically visible, and not just whether singularities
are visible. Looked at in this way, scalar collapse provides a serious
counterexample to censorship.

Similar critical behaviour has also been found in numerical studies
of collapse with other forms of matter. Axisymmetric collapse of
gravitational waves was shown to have a $\gamma$ of about 0.36, and
$\Delta \simeq 0.6$. For spherical collapse of radiation (perfect
fluid with equation of state $p=\rho/3$) the critical solution has
continuous self-similarity, and $\gamma$ of about 0.36. However it
has become clear now that the critical exponent $\gamma$ is not independent
of the choice of matter.
A study of collapse for a perfect fluid with an equation of state 
$p=k\rho$ shows that $\gamma$ depends on $k$. For a given form of
matter, there appears to be a unique $\gamma$, but the value changes as
the form of $T_{ik}$ is changed.

An important issue regarding the solutions with $p>p_{*}$ is the following.
These solutions are identified as black holes because of the presence of
an apparent horizon. However, current numerical studies do not probe the
singularity itself, and one cannot for now rule out the possibility that the
the solutions with $p>p_{*}$ fall into two classes: (a) those which have a 
Cauchy horizon lying outside the apparent horizon, and hence are naked
singularities, and (b) those which are black holes. There is actually some
evidence for this in the work of Brady~\cite{br}, and this aspect needs to be
investigated further.

A much more detailed discussion of scalar collapse and critical
phenomena can be found in other recent reviews~\cite{gun}.
 
\subsection{Spherical Collapse with General Form of Matter}

There is a certain degree of similarity in the collapse behaviour of
dust, fluids and scalar fields - in all cases some of the initial
data lead to black holes, while other data lead to
naked singularities. This would suggest an underlying pattern which is
probably characterized, not by the form of matter, but by some invariants
of the gravitational field. Hence investigations of collapse which put no 
restriction on $T_{ik}$ apart from an energy condition should prove useful.

An interesting attempt in this direction was made by
Dwivedi and Joshi~\cite{dj2}. They assumed a general $T^{i}_{k}$
obeying the weak energy condition, and also that the collapsing matter
forms a curvature singularity.
As we noted earlier, in the comoving coordinate system, matter is 
described by its energy density and the radial and tangential 
pressures. Along with these three functions, three functions describing
the metric enter a set of five Einstein equations, which are coupled
with an equation of state in order to close the system. The geodesic
equation for radial null geodesics is written in the limit of approach
to the singularity, and it is shown that the occurrence of a visible
singularity is equivalent to the occurrence of a positive real root for
the geodesic equation, suitably written. Since this equation depends on
free initial data, it follows that for a subset of the initial data there
will be positive real roots and the singularity will be visible.

\subsection{Null Geodesic Expansion and Cosmic Censorship}
 
A line of investigation which may prove useful for studying collapse of
a general form of matter is to examine the evolution of the expansion $\theta$
of a congruence of outgoing null geodesics. Some preliminary work has 
recently been done~\cite{the}. Consider first the case of the collapsing 
spherical
dust cloud. If a point on the cloud ends up as a covered singularity, then
$\theta$ at this point evolves to a negative value, as expected, starting
from its initial positive value. A naked singularity forms precisely in those
cases for which the initially positive $\theta$ continues to remain positive
all the way until singularity formation. We have given an argument suggesting
that this property of $\theta$ (i.e. its remaining positive throughout the
evolution for some initial data) is stable against small changes in the
equation of state.

Hence, if dust admits a naked singularity for some initial data, a naked
singularity will form also in the collapse of a fluid for which the ratio
of pressure to density is small but non-zero, provided one starts from the
same initial data. It may be possible to generalise these results, by using
the Raychaudhri equation to predict which initial conditions lead to a black
hole, and which ones to naked singularities. This is at present under
investigation.   

\subsection{Non-spherical Gravitational Collapse}

Amongst the very few studies of non-spherical collapse that have been
carried out so far is the numerical work of Shapiro and Teukolsky on
oblate and prolate collisionless spheroids. Since there really have been no
recent developments in this direction, I refer the reader
to the discussion of their work in Section 3 of my earlier 
review~\cite{tps}, and to the review by Wald~\cite{revi}.
 
A recent work on non-spherical perturbations of spherical collapse
deserves mention. Iguchi et al. have shown that the naked singularities
arising in spherical dust collapse are marginally stable against odd-parity
non-spherical gravitational wave perturbations~\cite{ig}.

Unlike the spherical case, very little is known about gravitational collapse
and cosmic censorship for non-spherical systems.

\subsection{Properties of Naked Singularities}

There are now sufficiently many known examples of naked singularities for one
to enquire about properties of such singularities. It may be that there
are well-defined laws of `naked singularity mechanics', just as there are
the laws of black hole mechanics (though there is no indication at the moment
that such a thing is true). At present there is only some scattered knowledge
about properties like curvature strength, stability of the Cauchy horizon and
redshift.

Examples of both weak curvature and strong curvature naked singularities
have been found~\cite{tps}. While spacetime cannot be extended through the 
latter kind
of singularity, it may possibly be extendible through a weak singularity.
For a discussion of extendibility see Clarke~\cite{cl}.

If the Cauchy horizon accompanying a naked singularity were to be unstable,
that could be evidence in favour of cosmic censorship. However, examples
of stable as well as unstable Cauchy horizons are known in classical
collapse. (See Penrose~\cite{revi} for some more discussion on Cauchy horizon
stability).

The redshift of the null rays emanating from a singularity can be shown
to be infinite, in the known examples, assuming that the standard redshift
definition can be used all the way up to a singularity. In this sense, naked
singularities are as black as black holes themselves. However, this does
not appear to be a good way to preserve censorship because ultimately quantum
effects near the naked singularity must be taken into account, and these will
serve to distinguish a black hole from a naked singularity.
  
It can also be shown in a straightforward way that any shell-focusing
naked singularities that might form in spherical collapse are necessarily
massless~\cite{lak}~\cite{coo}.

It can be said that if a naked singularity forms, its most significant  
property is that regions of extremely high curvature are exposed.
This will have observable consequences which will be essentially unaffected
by the other properties mentioned above. Hence these other properties can
only have secondary importance.

\subsection{Status of the Cosmic Censorship Hypothesis}

Until we learn something definite about non-spherical collapse, it is
not possible to conclude about the validity of the hypothesis. However
I would like to suggest, on the basis of what is known, that the hypothesis
is unlikely to be true in classical general relativity. We also note that
we are regarding a visible region of unbounded high curvature as a violation
of censorship, even if this visible region does not contain an actual
singularity.

The examples of naked singularities known in spherical collapse arise for
various forms of matter. This includes dust, perfect fluids, imperfect fluids
and scalar fields. There are also some general arguments suggesting the
occurrence of naked singularities for any form of matter satisfying the
weak energy condition (e.g. an existence proof, and the behaviour of null
geodesic expansion). None of this, taken by itself, constitutes a proof.
But, taken together, these arguments strongly suggest that
visible regions of unbounded curvature arise generically in spherical
gravitational collapse. Also, black holes arise generically in spherical
collapse.

Now, we know from the singularity theorems that the occurrence
of singularities in spherical collapse is stable against the introduction
of non-spherical perturbations. In view of this, it is very hard to see
why the naked singularities arising in spherical collapse should be unstable
against non-spherical perturbations, whereas the black holes forming in
spherical collapse should be stable against such perturbations.

It is only fair to say that different people have drawn widely different
conclusions about cosmic censorship from the currently known examples. Since
our viewpoint is that censorship possibly does not hold in classical 
relativity, we would like to ask next if naked singularities could actually
occur in nature, and if they do, what would they look like.

\section{Are there Naked Singularities in Nature?}

\subsection{Maybe No ...}

Even if general relativity were to generically admit naked singularity
solutions, it does not follow that these singularities actually occur
in nature. It could be that stars simply do not possess the initial
conditions necessary for formation of naked singularities.

Furthermore, there could actually be some principle, over and above general
relativity, which forbids naked singularities. This would be in the same 
spirit in which the advanced wave solutions of electrodynamics are forbidden.
We have recently pursued a line of thought wherein the second law of 
thermodynamics prohibits naked singularities~\cite{bar}.

Our idea can be deduced from Penrose's work on the second law of 
thermodynamics~\cite{pen}. As explained by Penrose, a fundamental 
understanding of the 
second law can be had only if we understand why the initial entropy of the
Universe is so low, compared to the maximal value it could have had. The 
matter, including radiation, was itself in a high entropy state because of 
the thermal equilibrium that prevailed soon after the Big Bang. Hence there 
must be an entropy associated with the gravitational field and this 
gravitational entropy must have been initially very low, so that the net 
entropy (matter plus gravity) becomes extremely small.

Such a gravitational entropy will have to be defined from the Riemann 
curvature. The Ricci part of the curvature diverges at a Friedmann Big
Bang singularity. Since we are interested in a low gravity entropy at the
Big Bang, it is plausible that this entropy is related to the Weyl part of
the curvature, which is zero at the Friedmann singularity. This has come to
be known as the Weyl Curvature Hypothesis: in order to have an explanation of
the second law, the Weyl curvature must be zero (or at least negligible
compared to the Ricci curvature) at the initial cosmological singularity.
It should be said though that a concrete mathematical relation between the 
Weyl curvature and gravitational entropy has not yet been found.

It is possible to regard a naked singularity forming in collapse as an 
`initial' singularity, because geodesics terminate in the past at the 
singularity. Hence it is reasonable to require that only those naked
singularities can occur which satisfy the Weyl hypothesis. That is, a suitable
quantity constructed from the Weyl curvature must go to zero as the naked
singularity is approached in the past along an outgoing geodesic. If a naked
singularity solution occurs in general relativity but violates the Weyl
hypothesis then its existence in nature is forbidden by the second law. Such
a naked singularity is a singularity with very high initial gravitational
entropy, contrary to what is expected for the second law to hold.

We tested the behaviour of the Weyl scalar in a few simple examples of 
spherical naked singularities and found it to diverge at the singularity, 
along outgoing geodesics.  It diverges as fast as the Ricci part, and hence 
violates the Weyl hypothesis. Since strong inhomogeneity tends to favour a 
high Weyl, and since it also favours naked singularities, it is likely that 
this divergence behaviour is generic to naked singularities. Thus naked
singularities may be anti-thermodynamic entities.

I do not know of any easy way out of this line of argument. The argument may
fail only if it turns out that there is actually no connection between
gravitational entropy and the Weyl curvature. 

\subsection{Maybe Yes ...}

It would of course be a much more interesting state of affairs if no principle
forbids naked singularities, and if they were to be found in nature. Hence we 
would like to enquire what the observational signatures of naked singularities
will be. It is nearly certain that naked singularities will not emit 
significantly through classical processes, because of the extremely large
redshifts. However, a quantum treatment of the matter and of gravity will be
unavoidable near the singularity, and these quantum effects will result in an
observable emission, in spite of the large classical redshift. In the absence
of a quantum theory of gravity, the best one can do is compute the quantum
particle creation in the classical gravitational field which becomes very
strong as the singularity is reached. This semiclassical treatment will in fact
suffice until the final Planck epoch prior to the singularity formation.

Thus, in effect one is asking what is the analog of Hawking radiation in the
case when a star collapses to a naked singularity. Answering this is not as
direct as the Hawking radiation calculation for a black hole, because of the
presence of a Cauchy horizon. Part of the future null infinity is exposed to 
the naked singularity and hence one cannot perform the usual expansion of 
matter field modes in the future. This prevents the usual Bogoliubov 
transformation and the standard particle creation calculation from being
carried out. This is one of the reasons why not much work has been done on this
important problem and computation is still in its infancy.

One way out is to compute the quantum expectation value of the stress energy
tensor - this can be calculated locally, and on future null infinity, in the
approach to the Cauchy horizon. The outgoing flux of radiation is a measure
of the emission from the naked singularity. In four dimensions an exact
calculation is not possible, but the outgoing flux in the geometric optics
approximation has been calculated~\cite{fp}
for dust shell-crossing singularities. In
this case the flux does not diverge in the approach to singularity formation.
We~\cite{va} have recently used the method of Ford and Parker to compute the 
flux of
a massless scalar field on the Cauchy horizon resulting from a shell-focusing
dust naked singularity. This time the flux diverges, suggesting that the 
back-reaction will avoid formation of the naked singularity. In an interesting
paper, Vaz and Witten~\cite{va2} have calculated the spectrum of this 
radiation, and shown
it to be very different from the black-body spectrum of Hawking radiation.

In two dimensions, as a result of the conformal anomaly, the outgoing flux
can be calculated exactly, without having to resort to the geometric optics
approximation. This was earlier done~\cite{his} for the null-dust (Vaidya) 
naked
singularity and repeated by us~\cite{bar2} for the dust (Tolman-Bondi) naked 
singularity.
In both cases the flux diverges on the Cauchy horizon. Similar features have
been found in studies~\cite{va3} of the quantum behaviour of naked 
singularities in some 
string-inspired gravity models. The back-reaction calculation in all the above
models is extremely hard to perform, as can be expected. But it is plausible
that essentially the back-reaction will remove the classical naked
singularity, without significantly affecting the flux emitted to infinity.

The observable signature of a naked singularity appears to be the burst of
radiation emitted as the Cauchy horizon is approached, and the characteristic 
non-thermal spectrum which this radiation possesses. This is to be contrasted 
with the slow evaporation of a quantum black hole via black-body radiation. It
would be important to generalise the above results to find the typical 
signatures of quantum naked singularities and to explore if there are any
astrophysical objects whose properties resemble those of a naked singularity.

\section*{Acknowledgments}

I would like to thank the organizers for inviting me to give a talk at this 
meeting.
It is a pleasure to thank Sukratu Barve, Srirang Deshingkar, I. H. Dwivedi, 
Sanjay Jhingan, Pankaj Joshi, Giulio Magli, Cenalo Vaz and Louis Witten
for many useful discussions.   
I acknowledge partial support of the {\it Junta Nacional de Investigac\~ao
Cient\'ifica e Tecnol\'ogica} (JNICT) Portugal, under contract number
CERN/S/FAE/1172/97.

\newpage

\centerline{\bf QUESTIONS}

\bigskip

\noindent{\bf C. Sivaram:} The third law of thermodynamics might have
something to do with, for instance, non-formation of a naked singularity from
an extremal black hole. You need an infinite number of steps to reduce $M$,
such that $M^{2}<Q^{2}$ or $a^{2}$, i.e. to destroy the event 
horizon.

\noindent{\bf T. P. Singh:} You maybe right, but perhaps we should be
a little cautious and note that there is no formal proof yet of the third
law of black hole mechanics within classical general relativity. It could
turn out that for such a proof to hold, cosmic censorship may have to be
assumed, as for the second law of black hole mechanics.

\medskip

\noindent{\bf C. Sivaram:} In a collapsing universe merger of black holes
would lead to an enormous increase of entropy, so that there is complete
time asymmetry with respect to the expanding phase. What would happen if
naked singularities are formed?

\smallskip

\noindent{\bf T. P. Singh:} If a naked singularity forms and the formation
of the horizon is prevented, then the huge entropy increase associated with the
horizon area will not take place. But it is not clear how much entropy is to
be associated with the naked singularity itself.  

\medskip

\noindent{\bf N. D. Hari Dass:} I do not understand your remark that
Choptuik's  result that arbitrarily small black holes can form is
not good for cosmic censorship, because howsoever small the black
hole is big enough to cover the singularity.

\smallskip

\noindent{\bf T. P. Singh:} I am avoiding making a physical distinction between
a singularity and a region of unbounded high curvature. The physical processes
that would result because of the formation of a naked singularity should be 
much the same as those resulting from the formation of a visible 
region of arbitrarily
high curvature but not in itself containing a singularity. Hence I think one
should regard cosmic censorship as the statement that visible regions of 
arbitrarily high curvature do not develop generically in collapse.
\medskip

\noindent{\bf Tariq Shahbaz:} How much energy would you expect to be
released in the formation of a naked singularity?

\smallskip

\noindent{\bf T. P. Singh:} At this stage, it is difficult to give a precise
answer to this question, and the answer will also be model dependent.
A few general remarks can be made. The energy release will come not only
because of the naked singularity, but also because of the high curvature 
regions surrounding it. Let there be a visible region of mass $M$, which has 
curvature high enough so that the associated curvature length scale is 
comparable to the Compton wavelength of some typical elementary particles.
Then one can expect almost the entire mass $M$ to be converted by pair creation
into the energy released during collapse.

\medskip

\noindent{\bf R. Misra:} How long will a naked singularity last?

\smallskip

\noindent{\bf T. P. Singh:} If we stay within the limits of classical
general relativity, spacetime comes to an end at a naked singularity, and
one cannot answer the above question. Also, from the point of view of a far
away observer, a classical naked singularity will take forever to form, because
of the infinite redshift. But quantum pair creation in the
collapsing object will be important, and then the naked singularity can be
expected to radiate itself away in a finite time (as seen by the distant
observer) as a result of the pair creation. We are trying to get estimates
of how much this time will be in some simple collapse models. Keeping in mind
the divergence of the radiated flux on the Cauchy horizon, one thing appears 
very likely - naked singularities are events (like explosions) and not 
objects with astronomical lifetimes.

\medskip

\noindent{\bf Chris Clarke:} How can you make any statement about the
Weyl Curvature Hypothesis for a singularity when viewed in the future,
given that in that region one is no longer in the domain of dependence
on the initial conditions?

\smallskip

\noindent{\bf T. P. Singh:} We are assuming an analytic continuation.

\medskip

\noindent{\bf Sukanta Bose:} Is it fair to compare a local and ``intensive''
quantity, namely the Weyl curvature, with a global and extensive quantity
such as gravitational entropy? Will not a given amount of matter, whether
in a mixed or pure state, affect the Weyl curvature identically?

\smallskip

\noindent{\bf T. P. Singh:} The connection between Weyl curvature and entropy,
if there is one, should be such that a non-local quantity constructed out
of the Weyl has properties of entropy. As regards the second question, if we
are considering classical matter, then we know that clumping (increase in
inhomogeneity) generally leads to increase in Weyl curvature. 

\medskip

\noindent{\bf J. Pasupathy:} Are the results of P. S. Joshi and T. P. Singh
on dust collapse analytical or numerical? $\rho_{1}<0$ seems physically
reasonable. Would you then say that naked singularities are more likely
than black holes? 

\smallskip

\noindent{\bf T. P. Singh:} The results are analytical. It is true that a naked
singularity is more likely than a black hole in the example I gave in the
talk (and mentioned here in the article). However, when the general initial
data for spherical dust collapse is examined, both black holes and naked
singularities result from a non-zero measure of the initial data set.


\section*{References}
\begin{thebibliography}{99}
\bibitem{rees}M. J. Rees in {\em Black Holes and Relativistic Stars} ed.
R. M. Wald (Chicago University Press, 1998), and articles in this volume.

\bibitem{pe}R. Penrose, \Journal {\it Rivista del Nuovo Cimento} {1} {252} 
{1969}.

\bibitem{revi}R. Penrose in {\em Black Holes and Relativistic Stars} 
ed. R. M. Wald (Chicago University Press, 1998) and in this volume; 
R. M. Wald, gr-qc/9710068; C. J. S. Clarke, \Journal{\CQG} {10} {1375}{1993}
and in this volume; P. S. Joshi in {\em Singularities, Black Holes and
Cosmic Censorship} ed. P. S. Joshi (IUCAA, Pune, 1997), gr-qc/9702036; 
P. S. Joshi, {\em Global Aspects in Gravitation and Cosmology} 
(Oxford, 1993).  

\bibitem{tps}T. P. Singh in {\em Classical and Quantum Aspects of Gravitation
and Cosmology} ed. G. Date and B. R. Iyer (Inst. of Math. Sc., Madras, 1996),
gr-qc/9606016.

\bibitem{dust} For detailed references on dust collapse see Ref. 2 above; two
recent papers are I. H. Dwivedi and P. S. Joshi, 
\Journal{\CQG}{14}{1223}{1997};
L. Herrera, A. Di Prisco, J. L. Hernandez-Pastora and N. O. Santos,
gr-qc/9711002.

\bibitem{vaidya} see Ref. 2.

\bibitem{op} A. Ori and T. Piran, \Journal{\PRD} { 42} {1068}{1990}

\bibitem{dj}  P. S. Joshi and I. H. Dwivedi, 
\Journal{\em Commun. Math. Phys.} {146} {333}{1992}.


\bibitem{lk} E. M. Lifshitz and I. M. Khalatnikov,  
\Journal{\em Soviet Physics JETP} {12} {108} {1961}. 


\bibitem{po}M. A. Podurets,
\Journal{\em Soviet Physics - Doklady} {11} {275} {1966}.

\bibitem{coo}F. I. Cooperstock, S. Jhingan, P. S. Joshi and T. P. Singh,
\Journal{\CQG}{14}{2195}{1997}.

\bibitem{si}P. Szekeres and V. Iyer, \Journal{ \PRD}{47} {4362} {1993}.


\bibitem{tp}G. Magli, \Journal{\CQG}{14}{1937}{1997}, gr-qc/9711082; 
T. P. Singh and L. Witten, \Journal{\CQG}{14}{3489}{1997};
S. Barve, T. P. Singh and L. Witten, in preparation.

\bibitem{ch} D. Christodoulou, \Journal{\em Commun. Math. Phys.} 
{105} {337} {1986}; {\bf 106}, 587 (1986);  {\bf 109}, 591 (1987);
{\bf 109}, 613 (1987); \Journal{\em Commun. Pure Appl. Math.} {XLIV}
{339} {1991}; {\bf XLVI}, {1131} {1993}; 
\Journal {\em Ann. Math.} {140} {607} {1994}. 


\bibitem{cho}M. W. Choptuik, \Journal {\em Phys. Rev. Lett.} {70} {9}{1993}.

\bibitem{br}P. R. Brady, \Journal{\PRD } {51} {4168} {1995}.


\bibitem{gun}M. W. Choptuik, gr-qc/9803075; C. Gundlach, gr-qc/9712084.

\bibitem{dj2} I. H. Dwivedi and P. S. Joshi, \Journal{\em Commun. Math. Phys.} 
{166} {117} {1994}. 


\bibitem{the}T. P. Singh, gr-qc/9711049.

\bibitem{ig}H. Iguchi, K. Nakao and T. Harada, Kyoto University preprint
KUNS 1475 (1997), to appear in {\em Phys. Rev.} D.

\bibitem{cl} C. J. S. Clarke, {\em Analysis of spacetime singularities}
(Cambridge University Press, 1993) and in this volume.

\bibitem{lak}K. Lake, \Journal{\em Phys. Rev. Lett.} {68} {3129}{1992}.

\bibitem{bar} S. Barve and T. P. Singh, \Journal{\em {Mod. Phys. Lett.} A}
{12} {2415}{1997}; gr-qc/9705060.

\bibitem{pen} R. Penrose in {\em Quantum Gravity 2} ed. C. J. Isham, 
R. Penrose and D. W. Sciama (Oxford, 1981).

\bibitem{fp} L. H. Ford and L. Parker \Journal{\PRD} {17}
{148}{1978}.

\bibitem{va} S. Barve, T. P. Singh, C. Vaz and L. Witten, gr-qc/9802035.

\bibitem{va2} C. Vaz and L. Witten, gr-qc/9804001.


\bibitem{his} W. A. Hiscock, L. G. Williams and D. M. Eardley,
\Journal{\PRD} {26} {751} {1982}.

\bibitem{bar2}S. Barve, T. P. Singh, C. Vaz and L. Witten, in preparation.

\bibitem{va3}C. Vaz and L. Witten, \Journal{\em {Nucl. Phys.} B}{487}{409}
{1997};
\Journal{\em {Phys. Lett.} B}{325}{27}{1994}, \Journal{\CQG}{13}{L59}{1996}.












\end{thebibliography}
\end{document}